\documentstyle[psfig]{article}
\begin{document}

\noindent{\Large{\bf Transverse effects on squeezing with atoms}}

\vspace{0.5cm}

{\sc Astrid Lambrecht, Jean-Michel Courty and Serge Reynaud} \\

{\it Laboratoire Kastler Brossel, UPMC, ENS, CNRS}

{\it Universit\'e Pierre et Marie Curie, case 74, F75252 Paris, France}\\

({\sc Journal de Physique II France} {\bf 6}, p. 1133 (1996))\\


\begin{abstract}
We evaluate the squeezing of a probe beam with a transverse Gaussian
structure interacting with an ensemble of two-level atoms in a cavity. We
use the linear input-output formalism where the effect of atoms is described
by susceptibility and noise functions, and show that the transverse
structure is accounted for by averaging these atomic functions over the
intensity profile. The results of the plane-wave and Gaussian-wave theories
are compared. We find that, when a large squeezing is predicted, the
prediction of the plane-wave model is not reliable outside the Kerr domain.
We give an estimate of the squeezing degradation due to the Gaussian
transverse structure.
\end{abstract}

\section{Introduction}

When an optical cavity containing a nonlinear medium approaches a
bistability turning point \cite{LugiatoBistab}, the fluctuations of the
incoming beam are squeezed \cite{LugiatoStrini}. Nearly perfect squeezing is
expected when the nonlinear medium is a lossless Kerr medium \cite
{Collett85,Reynaud89}. Squeezing was experimentally demonstrated at the
output of a bistable cavity containing an atomic beam \cite{Raizen87,Hope92}
or a cloud of cold atoms released from a magneto-optical trap \cite
{Lambrecht94}. A lot of theoretical work has been devoted to the case where
the nonlinear medium consists in two-level atoms \cite
{Reid85,Reid86a,Reid86b,Carmichael,Lugiato86,Orozco87,Castelli,Reid88,Hilico}.

Most theoretical treatments have simplified the description of the
field-matter interaction by modelizing the laser beam as a plane-wave. It is
however clear that the description of the bistable behaviour of the cavity
has to be modified to account for the transverse Gaussian profile of the
laser beam \cite{Drummond81,Ballagh,LugiatoGW,Rosenberger}. This is also
true for the predictions concerning field fluctuations at the output of the
cavity, as shown in references \cite{Xiao87a,Xiao87b,Hope90}. These papers
contain a derivation of squeezing spectra in optical bistability accounting
for the transverse Gaussian structure of the laser beam. For particular sets
of parameters, the spectra obtained in Gaussian-wave calculations are
compared by the authors with the predictions of the plane-wave model. In the
good-cavity limit where the atomic damping occurs on a much shorter time
than the intracavity field damping and where the atomic variables can be
adiabatically eliminated, Xiao et al \cite{Xiao87a} conclude that the
plane-wave calculations may be trusted in either the weak-field or
dispersive limits. In a complete treatment without adiabatic elimination 
\cite{Hope90}, Hope et al relate squeezing degradation due to transverse
structure to the ratio of atomic and cavity damping times.

In the present paper, we quantitatively delineate the regions in the
parameter space where the predictions of the plane-wave and Gaussian-wave
calculations coincide, and therefore estimate the reliability of plane-wave
model. We obtain criteria for reliability of plane-wave model which are
quite different from those of references \cite{Xiao87a,Xiao87b,Hope90}.

We will use the linear input-output formalism generalized to incorporate the
treatment of atomic fluctuations \cite{Hilico}. This formalism is based upon
a linearisation of the fluctuations around the working point of the bistable
cavity, as the quantum stochastic methods \cite{CollettGardiner}, and it
provides the same spectra as the latter when the same system is studied with
the same simplifying assumptions \cite{Reynaud89OC}. In contrast with the
quantum stochastic methods, it provides an intuitive understanding of the
noise processing occuring in the bistable cavity. The squeezing process
through the parametric transformation of vacuum fluctuations going into the
intracavity mode is described in terms of atomic susceptibility functions,
while the squeezing degradation through the addition of fluctuations by the
atoms is characterized by atomic noise functions.

We will show in the following that the effect of Gaussian profile upon noise
processing appears quite naturally as a spatial averaging of these
susceptibility and noise functions. This will allow us to obtain simple
techniques for describing the influence of transverse structure. It is worth
stressing that the property of averaging of the susceptibility and noise
functions is not postulated a priori, but is rather a consequence of the
simple assumptions made in the treatment of Gaussian profile: we assume that
the higher-order transverse modes are either far from resonance with the
cavity, or of such a high order that they are not efficiently coupled to the
fundamental mode. It is thus possible to project any field radiated by the
atoms upon this mode, and to obtain the averaging property as a consequence
of this projection.

Using this technique, we will first recover the well-known result \cite
{Drummond81,Ballagh,LugiatoGW,Rosenberger} that the mean fields evaluated
with the plane-wave and Gaussian-wave calculations coincide in the
low-saturation limit where the population of the atomic excited state
remains small. This implies that the bistability curves relating the input
and intracavity intensities also coincide provided that these intensities
are properly defined. We will then evaluate the atomic susceptibility and
noise functions, and deduce that the conditions for coincidence of
plane-wave and Gaussian-wave calculations are much more restrictive in this
case: the limit of a small population in the excited state is not sufficient
for ensuring the coincidence for atomic noise functions. In particular, we
will conclude that the noise functions evaluated in the plane-wave model are
not reliable when large squeezing is predicted with parameters lying outside
the Kerr domain. We will finally give a quantitative estimate for the
degradation of squeezing due to transverse structure.

\section{Model and assumptions}

We consider a single ended ring cavity containing a nonlinear medium. A
probe beam is entering the cavity through the coupling mirror, interacting
with the medium and leaving the cavity through the same mirror. To study the
influence of this device on the quantum fluctuations of the probe beam, we
use a semi-classical input-output formalism, in which the field fluctuations
are considered as classical stochastic variables, which are driven by the
quantum fluctuations of the incoming fields at the laser frequency $\omega _L
$. The linear input-output relations give the quantum fluctuations of the
output field as a function of those of the input field, in cases when they
are small compared to the mean field-values, an assumption which is
particularly well adapted for high photon numbers, i.e. laser beams in a
resonator. The method is described in detail in \cite{Hilico}.

The intracavity field is defined as Heisenberg operator in the frame
rotating with respect to the laser frequency $\omega _L$:

\begin{equation}
e^{-i\omega _Lt}E\left( {\bf r},t\right) +e^{i\omega _Lt}E\left( {\bf r%
},t\right) ^{\dagger }  \label{FieldOperators}
\end{equation}
In order to account for the spatial beam structure, the components $E\left( 
{\bf r},t\right) $ and $E\left( {\bf r},t\right) ^{\dagger }$ are
written as products of a Gaussian mode function by a time-dependent mode
amplitude (${\bf r}$ stands for the three-dimensional position $\left(
x,y,z\right) $):

\begin{eqnarray}
E\left( {\bf r},t\right) &=&\sqrt{\frac{\hbar \omega _L}{2\varepsilon _0c}%
}A\left( t\right) u\left( {\bf r}\right)  \label{E} \\
E\left( {\bf r},t\right) ^{\dagger } &=&\sqrt{\frac{\hbar \omega _L}{%
2\varepsilon _0c}}A\left( t\right) ^{\dagger }u\left( {\bf r}\right) ^{*}
\label{Ecross}
\end{eqnarray}
This definition is such that $\left\langle A\left( t\right) ^{\dagger
}A\left( t\right) \right\rangle $ is the number of photons going through a
beam section in the unit time.

The normalized Gaussian mode functions are written for a propagation along
the $x$ direction:

\begin{equation}
u\left( {\bf r}\right) =\sqrt{\frac{f\left( {\bf r}\right) }{S\left(
x\right) }}\exp \left[ -i\varphi \left( {\bf r}\right) \right] 
\label{GaussianMode}
\end{equation}
$f(r)$ is the Gaussian intensity normalized to unity on the beam axis, $S$
the effective beam section and $\varphi (r)$ the phase:

\begin{eqnarray}
f\left( {\bf r}\right) &=&\exp \left[ \frac{-2\left( y^2+z^2\right) }{%
w^2\left( x\right) }\right] \\
S\left( x\right) &=&\frac{\pi w^2\left( x\right) }2 \\
\varphi \left( {\bf r}\right) &=&-\frac{2\pi x}\lambda +\arctan
\left( \frac x{l_R}\right) -\frac \pi \lambda \frac{\left( y^2+z^2\right) x}{%
x^2+l_R^2}
\end{eqnarray}
$w\left( x\right) $ denotes the position-dependent beam size, $w_0$ the beam
waist, $l_R$ the Rayleigh divergence length and $\lambda $ the laser
wavelength:

\begin{eqnarray}
w^2\left( x\right) &=&w_0^2\left( 1+\frac{x^2}{l_R^2}\right) \\
l_R &=&\frac{\pi w_0^2}\lambda
\end{eqnarray}

Since we suppose the cavity to be perfectly mode matched to the laser beam,
the same transverse structure holds for the input field and for the
intracavity field before its interaction with the atoms. This structure is
modified for the output field however, through the non linear interaction
with the atoms on one hand, and through spontaneous emission of light in all
possible directions by the atoms on the other hand. Both effects couple the
fundamental Gaussian mode to all the other cavity modes. We will suppose
here that the field radiated by the atoms into the other modes has an effect
on intracavity fields much smaller than the field radiated into the
fundamental Gaussian mode. This may be due either to the fact that the
fundamental Gaussian mode is close to a cavity resonance while the other
modes at the same frequency are far from any cavity resonance, or that they
have such a high order that their coupling with the fundamental mode is
negligeable. We will also consider that the detection setup is perfectly
matched to the Gaussian mode.

As a consequence, we can write the input and output fields $E^{{\rm {in}}}(%
{\bf r},t)$ and $E^{{\rm {out}}}({\bf r},t)$ in the same manner as
the intracavity field $E({\bf r},t)$:

\begin{eqnarray}
E^{{\rm {in}}}\left( {\bf r,}t\right) &=&\sqrt{\frac{\hbar \omega _L}{%
2\varepsilon _0c}}A^{{\rm {in}}}\left( t\right) u\left( {\bf r}\right) \\
E^{{\rm {out}}}\left( {\bf r,}t\right) &=&\sqrt{\frac{\hbar \omega _L}{%
2\varepsilon _0c}}A^{{\rm {out}}}\left( t\right) u\left( {\bf r}\right)
\end{eqnarray}
Clearly the reflection-transmission equations for the fields on the coupling
mirror will be the same for the field amplitudes as for the plane-wave
fields in \cite{Hilico}. In contrast, the effect of the atoms on these field
amplitudes will be modified, due to the non-linearity of the interaction and
to the Gaussian variation of laser intensity. In order to evaluate the
modification of the mode amplitudes due to the atomic medium, we will
disregard the fields radiated by the atoms into the higher-order modes and
obtain these modifications by projecting the radiated field onto the
fundamental mode. This will be the key simplifying assumption in the
forthcoming calculations concerning the mean fields as well as the field
fluctuations.

We will also use two simplifying hypothesis. First, we will consider that
the transverse size $w$ of the probe beam is much smaller than the size of
the atomic medium, so that the dependence of the atomic density versus the
transverse variables $y$ and $z$ may be disregarded. This density $\rho
\left( {\bf r}\right) $ will therefore be replaced by the on-axis atomic
density $\rho _{{\rm m}}$:

\begin{equation}
\rho _{{\rm m}}(x)=\rho (x,y=0,z=0)
\end{equation}
Note that the longitudinal dependence of the atomic density will have to be
accounted for, since we consider that the length of the interaction zone is
limited by the size of the atomic medium. Second, we will assume that the
Rayleigh divergence length $l_R$ is much larger than this size, so that the $%
x$-dependence of the laser beam size $w$ can be ignored. However, we will
not suppose that the atomic medium is centered on the cavity waist.

\section{Description of mean fields in a Gaussian-wave theory}

The atomic medium is considered to be a homogeneously broadened system of
two-level atoms with a resonance frequency $\omega _0$, which is placed
inside an optical cavity and driven by a coherent mean field of frequency $%
\omega _L$ close to a cavity mode $\omega _C$. Atom-laser detuning $\delta $
and cavity-laser detuning $\phi $ are respectively normalised by the atomic
dipole decay rate $\gamma $ and the cavity field decay rate $\kappa $:

\begin{eqnarray}
\delta &=&\frac{\omega _0-\omega _L}\gamma \\
\phi &=&\frac{\omega _C-\omega _L}\kappa
\end{eqnarray}

It follows from the discussions of the previous section that the evolution
of the intracavity Gaussian mode amplitude during one cavity round trip time
is described by the same differential equation as in the plane-wave model:

\begin{equation}
\tau _C\partial _tA(t)=-\kappa \tau _C\left( 1+i\phi \right) A(t)+\sqrt{%
2\kappa \tau _C}A^{{\rm in}}(t)+dA(t)  \label{EquaDiff}
\end{equation}
However, the modification $dA$ of the intracavity Gaussian mode amplitude
due to one passage through the atomic medium has not the same expression as
in the plane-wave model, as a consequence of the nonlinearity of atoms-field
interaction. In the present section, we evaluate its mean value and then
deduce the bistability curve.

The effect of an optically thin atomic layer of length $dx$ may be described
by a local modification $dE\left( {\bf r}\right)$ of the field:

\begin{equation}
E\left( {\bf r}\right) \rightarrow E\left( {\bf r}\right) +dE\left( 
{\bf r}\right)
\end{equation}
with \cite{Heidmann85}:

\begin{equation}
dE\left( {\bf r}\right) =-\frac{3\lambda ^2}{4\pi }\rho \left( {\bf r}%
\right) \alpha \left( {\bf r}\right) E\left( {\bf r}\right) dx
\end{equation}
$\alpha \left( {\bf r}\right) $ is the atomic polarizability derived from
the solution of the optical Bloch equations \cite{Hilico} and measured as a
dimensionless number:

\begin{equation}
\alpha \left( {\bf r}\right) =\frac{\alpha _l}{1+2s\left( {\bf r}%
\right) }
\end{equation}
$\alpha _l$ is the dimensionless linear polarizability of the atomic dipole:

\begin{equation}
\alpha _l=\frac 1{1+i\delta }
\end{equation}
$s\left( {\bf r}\right) $ is the dimensionless saturation parameter
proportional to the local laser intensity:

\begin{equation}
s\left( {\bf r}\right) =\frac{\left| \beta \left( {\bf r}\right)
\right| ^2}{1+\delta ^2}  \label{sat}
\end{equation}
$\beta \left( {\bf r}\right) $ is the dimensionless field parameter, that
is precisely the Rabi frequency normalized to the decay rate of the dipole:

\begin{equation}
\beta \left( {\bf r}\right) =\frac{d_0E\left( {\bf r}\right) }{\hbar
\gamma }=\sqrt{\frac{3\lambda ^2}{4\pi }}\frac A{\sqrt{\gamma }}u\left( 
{\bf r}\right)  \label{beta}
\end{equation}
where $d_0$ is the matrix element of the atomic dipole. These relations may
also be written in terms of the saturation parameter $s_{{\rm m}}$ and of
the field parameter $\beta _{{\rm m}}$ evaluated on the beam axis:

\begin{eqnarray}
s\left( {\bf r}\right) &=&s_{{\rm m}}f\left( {\bf r}\right) \\
s_{{\rm m}} &=&\frac{\left| \beta _{{\rm m}}\right| ^2}{1+\delta ^2}
\label{satm} \\
\beta _{{\rm m}} &=&\sqrt{\frac{3\lambda ^2}{4\pi S}}\frac A{\sqrt{\gamma }}
\label{betam}
\end{eqnarray}

Note that we may forget the $x$-dependence of the on-axis saturation
parameter $s_{{\rm m}}$ as a consequence of the assumption of a large
Rayleigh divergence length. The more general case of an arbitrary Rayleigh
divergence length might as well be studied \cite{AtomicNumber}. Using the
transverse monomode approximation discussed above, we project $dE$ onto the
Gaussian mode, and write the following equation for the modification $dA$ of
the mode amplitude, summing up over the layers inside the atomic medium:

\begin{equation}
\frac{dA}A=-\frac{3\lambda ^2}{4\pi S}\alpha _ln_s
\end{equation}
$n_{{\rm s}}$ is the effective number of atoms in the probe beam, taking
into account the effect of the saturation:

\begin{equation}
n_{{\rm s}}=\int \frac{f\left( {\bf r}\right) }{1+2s_{{\rm m}}f\left( 
{\bf r}\right) }\rho \left( {\bf r}\right) d{\bf r}
\end{equation}

In order to evaluate $n_{{\rm s}}$, we expand it in powers of the
saturation parameter:

\begin{eqnarray}
n_{{\rm s}} &=&\displaystyle{\sum _{k=0}^{\infty}}\left( -2s_{{\rm %
m}}\right) ^kn^{\left( 1+k\right) }  \label{powerexpansionN} \\
n^{\left( 1+k\right) } &=&\int f\left( {\bf r}\right) ^{1+k}\rho \left( 
{\bf r}\right) d{\bf r}
\end{eqnarray}
$n^{\left( 1\right) }$, that we will denote simply $n$, is the effective
number of atoms in the probe beam, defined in the unsaturated regime:

\begin{equation}
n=\int f\left( {\bf r}\right) \rho \left( {\bf r}\right) d{\bf r}
\end{equation}
The next term $n^{\left( 2\right) }$ corresponds to the lowest-order
saturation effect, that is precisely the Kerr nonlinearity of the atomic
medium. The other terms represent higher-order saturation effects. Due to
the Gaussian shape of the function $f\left( {\bf r}\right) $, the
function $f\left( {\bf r}\right) ^{{1+k}}$ has the same expression as $%
f\left( {\bf r}\right) $ with a modified value of the beam size
parameter, precisely with $w^{{2}}$ replaced by $w^{{2}}/\left( 1+k\right) $%
; $n^{\left( 1+k\right) }$ has therefore the same expression as $n$ with $w^{%
{2}}$ replaced by $w^{{2}}/\left( 1+k\right) $.

We now use the assumption that the transverse size $w$ of the probe beam is
much smaller than the size of the atomic medium. The unsaturated effective
number $n$ has thus a simple expression in terms of the on-axis atomic
density:

\begin{equation}
n=\frac{\pi w^{{2}}}2\int \rho _{{\rm m}}\left( x\right) dx
\end{equation}
The higher-order integrals $n^{\left( 1+k\right) }$ may therefore be written
as:

\begin{equation}
n^{\left( 1+k\right) }=\frac n{1+k}
\end{equation}
The saturated effective number $n_{{\rm s}}$ can then be related to the
unsaturated effective number $n$ through an analytical expression:

\begin{equation}
n_{{\rm s}}=nF\left( s_{{\rm m}}\right)
\end{equation}
\begin{equation}
F\left( s_{{\rm m}}\right) =1-\frac{2s_{{\rm m}}}2+\frac{\left( 2s_{{\rm m%
}}\right) ^2}3-\frac{\left( 2s_{{\rm m}}\right) ^3}4+..=\frac 1{2s_{{\rm m}%
}}\log \left( 1+2s_{{\rm m}}\right)  \label{Fgauss}
\end{equation}
This may be compared with the corresponding equation in the plane-wave model:

\begin{equation}
n_{{\rm s}}=nF_{{\rm pw}}\left( s\right)
\end{equation}
\begin{equation}
F_{{\rm pw}}\left( s\right) =1-2s+\left( 2s\right) ^2-\left( 2s\right)
^3+..=\frac 1{1+2s}  \label{Fpw}
\end{equation}

The two expressions nearly coincide in the low saturation regime, where only
the linear index and the Kerr term are appreciable, provided that an
effective value $s_{{\rm e}}$ of the saturation parameter is chosen in the
plane-wave model such that:

\begin{equation}
s_{{\rm m}}=2s_{{\rm e}}  \label{CompGaussPW}
\end{equation}
In more physical terms, this condition means that, in the Kerr limit, the
plane-wave and Gaussian results have to be compared for equal mean
intensities, i.e. equal intensities after averaging over the Gaussian
profile. As soon as the saturation parameter is appreciable in contrast ($%
s\approx 1$ or $s>1$), the two expressions differ. More strikingly, the
expression (\ref{Fgauss}) which accounts for transverse mode profile does
not tend towards the plane-wave expression (\ref{Fpw}) when the beam size
parameter grows. In this regime also, valid conclusions about the effect of
gaussian beam structure may be reached only after a criterium has been
chosen to compare plane-wave and Gaussian-beam spectra. In the following, we
will always compare spectra with equation (\ref{CompGaussPW}) obeyed.

The modification $dA$ of the intracavity Gaussian mode amplitude due to one
passage in the atomic medium may now be written:

\begin{equation}
\frac{dA}A=-\frac{3\lambda ^2}{4\pi S}\frac{nF\left( s_{{\rm m}}\right) }{%
1+i\delta }
\end{equation}
We easily obtain the steady state solution of the differential equation (\ref
{EquaDiff}) for the intracavity amplitude $A$:

\begin{equation}
\sqrt{2\kappa \tau _C}A^{{\rm in}}=\kappa \tau _C\left( 1+i\phi \right) A-dA
\end{equation}
which may be written as:

\begin{equation}
\frac 2{\sqrt{T}}\beta ^{{\rm in}}=\beta _{{\rm m}}\left[ \left( 1+i\phi
\right) +2C\frac{F\left( s_{{\rm m}}\right) }{1+i\delta }\right]
\end{equation}
$\beta _{{\rm m}}$ denotes the dimensionless parameter for the on-axis
intracavity field (cf. equation(\ref{betam})) and $\beta ^{{\rm in}}$ a
dimensionless parameter for the input field defined in an analogous manner:

\begin{equation}
\beta ^{{\rm in}}=\sqrt{\frac{3\lambda ^2}{4\pi S}}\frac{A^{{\rm in}}}{%
\sqrt{\gamma }}
\end{equation}
$C$ is the cooperativity parameter, which is proportional to the effective
number $n$ of atoms present in the probe beam:

\begin{equation}
C=\frac{3\lambda ^2}{4\pi S}\frac nT
\end{equation}
$T$ is the intensity transmission of the coupling mirror:

\begin{equation}
T=2\kappa \tau _C
\end{equation}
We then derive the relation between the input intensity and the intracavity
intensity:

\begin{equation}
\frac 4T\left| \beta ^{{\rm in}}\right| ^2=\left| \beta _{{\rm m}}\right|
^2\left| \left( 1+i\phi \right) +2C\frac{F\left( s_{{\rm m}}\right) }{%
1+i\delta }\right| ^2
\end{equation}
or:

\begin{equation}
Y=X\left[ \left( 1+CF^{\prime }\left( X\right) \right) ^2+\left( \phi
-\delta CF^{\prime }\left( X\right) \right) ^2\right]  \label{bistab}
\end{equation}
where we have replaced the field parameters by normalized intensities $X$
and $Y$ defined from the intensities averaged over the Gaussian profile:

\begin{equation}
Y=\frac 2T\left| \beta ^{{\rm in}}\right| ^2  \label{Y}
\end{equation}
\begin{equation}
X=\frac 12\left| \beta _{{\rm m}}\right| ^2=\frac 12s_{{\rm m}}\left(
1+\delta ^2\right)  \label{X}
\end{equation}
\begin{equation}
F^{\prime }\left( X\right) =\frac{2F\left( s_{{\rm m}}\right) }{1+\delta ^2}%
=\frac 1{2X}\log \left( 1+\frac{4X}{1+\delta ^2}\right)  \label{G}
\end{equation}
Equation (\ref{bistab}) is identical with results of \cite
{Drummond81,Ballagh,LugiatoGW,Rosenberger,Xiao87a,Hope90}. It must be
compared with the corresponding equation in the plane-wave model:

\begin{equation}
Y=X\left[ \left( 1+CF_{{\rm pw}}^{\prime }\left( X\right) \right) ^2+\left(
\phi -\delta CF_{{\rm pw}}^{\prime }\left( X\right) \right) ^2\right]
\label{bistabPW}
\end{equation}
where the normalized intensities $X$ and $Y$ are now defined from the
position-independent intensities in the plane-wave model:

\begin{equation}
Y=\frac 4T\left| \beta ^{{\rm in}}\right| ^2  \label{Ypw}
\end{equation}
\begin{equation}
X=\left| \beta _{{\rm e}}\right| ^2=s_{{\rm e}}\left( 1+\delta ^2\right)
\label{Xpw}
\end{equation}
\begin{equation}
F_{{\rm pw}}^{\prime }\left( X\right) =\frac{2F_{{\rm pw}}\left( s_{{\rm e%
}}\right) }{1+\delta ^2}=\frac 2{1+\delta ^2+2X}  \label{Gpw}
\end{equation}

As long as the saturation parameter remains small ($X\ll 1+\delta ^2$), the
bistability curves (\ref{bistab}) and (\ref{bistabPW}) are identical. Note
that this regime corresponds to a low-saturation limit where the mean
population of the atomic excited state defined in the plane-wave model
remains small:

\begin{equation}
\frac X{1+\delta ^2+2X}\ll 1  \label{lowsat}
\end{equation}
In the following, we will focus the attention on this regime where the
predictions of the plane-wave and Gaussian-wave theories coincide for the
mean fields. As already emphasized in the introduction, this does not imply
that the predictions will also coincide for the fluctuations.

Before coming to the study of fluctuations, we want to give conditions for
obtaining a large squeezing. A first condition is that a bistability turning
point is approached (\cite{Reynaud89} and references therein). This
condition can be formulated in more quantitative terms by writing the slope
of the bistability curve:

\begin{equation}
\frac{dY}{dX}=\left( 1+a\right) ^2+\left( \phi -a\delta \right) ^2-\zeta ^2
\label{slope}
\end{equation}
where:

\begin{eqnarray}
a &=&\frac{2C\left( 1+\delta ^2\right) }{\left( 1+\delta ^2+2X\right) ^2}
\label{defalpha} \\
\zeta &=&\frac{4CX\sqrt{1+\delta ^2}}{\left( 1+\delta ^2+2X\right) ^2}
\label{defzeta}
\end{eqnarray}
The coefficient $a$ describes the damping of intracavity field due to atomic
absorption, normalized to the damping due to the coupling mirror while the
coefficient $\zeta $ represents the normalized strength of the Kerr coupling
due to atomic non-linearity. The presence of a turning point ($dY/dX=0$)
requires that the Kerr coupling is large enough so that third term in the
right-hand side of equation (\ref{slope}) is larger than the first one:

\begin{equation}
\zeta \geq 1+a  \label{condzeta}
\end{equation}
and then that the detuning $\phi $ of empty cavity is chosen such as to
reach one of the two turning points:

\begin{equation}
\phi -a\delta =\pm \sqrt{\zeta ^2-\left( 1+a\right) ^2}  \label{condphi}
\end{equation}
The second condition for getting a large squeezing is that bistability is
mainly in the dispersive regime, which means that the effect of atomic
absorption on the field remains small:

\begin{equation}
a\ll 1  \label{condalpha}
\end{equation}
If this condition is not satisfied, atoms efficiently couple vacuum
fluctuations coming from the empty modes responsible for spontaneous
emission into the intracavity field fluctuations, leading to a degradation
of the expected squeezing.

In the low-saturation regime ($X\ll 1+\delta ^2$), conditions (\ref
{condalpha},\ref{condzeta}) imply a large atom-laser detuning ($\delta ^2\gg
1$) and they may be rewritten:

\begin{equation}
\delta \leq 2C\ll \delta ^2  \label{conditionC}
\end{equation}
\begin{equation}
\frac{\delta ^3}{4C}\leq X  \label{conditionX}
\end{equation}
In the following, we will consider that these conditions delineate the
region where a large squeezing may be obtained. It is in particular worth
noting that the intensity parameter $X$ lies in the domain $\delta \ll X\ll
\delta ^2$.

\section{Description of fluctuations in a Gaussian-wave theory}

We are now going to study the quantum fluctuations of the field. We will
closely follow the reference \cite{Hilico} where quantum fluctuations were
derived from a linear input-output theory with a plane-wave model. As in the
previous analysis of mean values, the reflection-transmission equations for
the field fluctuations on the coupling mirror are the same as in the
plane-wave model. In contrast, the effect of the atoms on these fluctuations
is modified, due to the non-linearity of the interaction and to the Gaussian
profile of laser intensity.

In order to discuss the transformation of field fluctuations, we now
introduce a new notation for the positive-frequency and negative-frequency
components $E\left( {\bf r},t\right) $ and $E^{\dagger }\left( {\bf r}%
,t\right) $ of the field operator (see equation (\ref{FieldOperators}))
which are coupled by the nonlinear interaction. They will be considered as
the components ${\cal E}_\alpha $ of a two-fold column matrix ${\cal E}
$:

\begin{equation}
{\cal E}\left( {\bf r,}t\right) =\left[ 
\begin{tabular}{l}
$E\left( {\bf r,}t\right) $ \\ 
$E^{\dagger }\left( {\bf r,}t\right) $%
\end{tabular}
\right]
\end{equation}
We will also denote ${\cal E}^\alpha ={\cal E}_\alpha ^{\dagger }$ the
components of the adjoint line matrix ${\cal E}^{\dagger }$.

The field components are written in terms of mode amplitudes and Gaussian
mode functions:

\begin{equation}
{\cal E}_\alpha \left( {\bf r},t\right) =\sqrt{\frac{\hbar \omega _L}{%
2\varepsilon _0c}}{\cal A}_\alpha \left( t\right) {\cal U}_\alpha
\left( {\bf r}\right)
\end{equation}
The two components ${\cal U}_\alpha \left( {\bf r}\right) $ coincide
with the functions $u\left( {\bf r}\right) $ and $u\left( {\bf r}%
\right) ^{*}$ defined previously in equation (\ref{E}-\ref{GaussianMode}):

\begin{eqnarray}
{\cal U}_\alpha \left( {\bf r}\right) &=&\sqrt{\frac{f\left( {\bf r}%
\right) }{S\left( x\right) }}\exp \left( -i\varepsilon _\alpha \varphi
\left( {\bf r}\right) \right) \\
\varepsilon _\alpha &=&\left( -1\right) ^{1+\alpha }
\end{eqnarray}
The same notation will be used also for the positive-frequency and
negative-frequency mode amplitudes:

\begin{equation}
{\cal A}\left( t\right) =\left[ 
\begin{tabular}{l}
$A\left( t\right) $ \\ 
$A^{\dagger }\left( t\right) $%
\end{tabular}
\right]
\end{equation}
as well as for the components of the atomic dipole measured as a
dimensionless number:

\begin{equation}
{\cal S}\left( t\right) =\left[ 
\begin{tabular}{l}
$S\left( t\right) $ \\ 
$S^{\dagger }\left( t\right) $%
\end{tabular}
\right]
\end{equation}
($S\left( t\right) $ and $S^{\dagger }\left( t\right) $ have the same
definition as in \cite{Hilico}; they will be denoted ${\cal S}_\alpha
\left( t\right) $ in the following).

As for the mean values, the effect of an optically thin atomic layer of
length $dx$ may be described by a local modification of the field
fluctuations:

\begin{equation}
\delta {\cal E}_\alpha \left( {\bf r},t\right) \rightarrow \delta 
{\cal E}_\alpha \left( {\bf r},t\right) +d\left( \delta {\cal E}%
_\alpha \left( {\bf r},t\right) \right)
\end{equation}
We have introduced here a generic notation for the fluctuations $\delta O$
of an operator $O$:

\begin{equation}
\delta O=O-\left\langle O\right\rangle
\end{equation}
The field modification $d\left( \delta {\cal E}_\alpha \right) $ may be
separated into two parts \cite{Hilico} representing respectively the linear
response $d\left( \delta {\cal E}_\alpha ^{{\rm lr}}\right) $ of the
atomic dipoles to the intracavity field fluctuations and the spontaneous
emission noise $d\left( \delta {\cal E}_\alpha ^{{\rm se}}\right) $,
that is the linear response of atomic dipoles to the vacuum fluctuations in
the field modes coupled to the atoms which are different from the
intracavity modes \cite{Courty91,Courty92}:

\begin{equation}
d\left( \delta {\cal E}_\alpha \left( {\bf r},t\right) \right)
=d\left( \delta {\cal E}_\alpha ^{{\rm lr}}\left( {\bf r},t\right)
\right) +d\left( \delta {\cal E}_\alpha ^{{\rm se}}\left( {\bf r}%
,t\right) \right)  \label{deltaE}
\end{equation}
The spontaneous emission term is uncorrelated with the intracavity field
fluctuations. Note that we disregard the effect of the higher-order
intracavity modes, in consistency with our assumptions on the treatment of
transverse mode structure. We now evaluate these two terms.

The first term may be written in terms of linear susceptibility functions $%
\chi _\alpha ^\beta $:

\begin{equation}
d\left( \delta {\cal E}_\alpha ^{{\rm lr}}\left( {\bf r},t\right)
\right) =-\frac{3\lambda ^2}{4\pi }\rho \left( {\bf r}\right) dx\int 
\displaystyle{\sum_{\beta}}\chi _\alpha ^\beta \left( {\bf r},\tau \right)
\delta {\cal E}_\beta \left( {\bf r},t-\tau \right) d\tau
\label{deltaElr}
\end{equation}
The susceptibility function $\chi _\alpha ^\beta \left( {\bf r},\tau
\right) $ is deduced from commutators of the atomic dipoles evaluated at
different times:

\begin{equation}
\chi _\alpha ^\beta \left( {\bf r},\tau \right) =i\Theta \left( \tau
\right) \left\langle \left[ {\cal S}_\alpha \left( \tau \right) ,{\cal %
S}^\beta \left( 0\right) \right] \right\rangle _{{\bf r}}  \label{khit}
\end{equation}
$\Theta \left( \tau \right) $ is the Heaviside function. The symbol $%
\left\langle ...\right\rangle _{{\bf r}}$ means that the correlation
functions are evaluated from the solution of the optical Bloch equations for
an atom located at point ${\bf r}$. As usually, the optical Bloch
equations are solved in the frequency domain rather than in the time domain.
We will give below the solution for the Fourier transforms $\chi _\alpha
^\beta \left( {\bf r},\left[ \omega \right] \right) $ of $\chi _\alpha
^\beta \left( {\bf r},\tau \right) $ defined according to the general
prescription:

\[
f\left( t\right) =\int \frac{d\omega }{2\pi }e^{-i\omega t}f\left[ \omega
\right] 
\]
The field modification $d\left( \delta {\cal E}_\alpha ^{{\rm lr}%
}\right) $ will then be written in the frequency domain:

\begin{equation}
d\left( \delta {\cal E}_\alpha ^{{\rm lr}}\left( {\bf r},\left[
\omega \right] \right) \right) =-\frac{3\lambda ^2}{4\pi }\rho \left( 
{\bf r}\right) dx\displaystyle{\sum_{\beta} }\chi _\alpha ^\beta \left( 
{\bf r},\left[ \omega \right] \right) \delta {\cal E}_\beta \left( 
{\bf r},\left[ \omega \right] \right)
\end{equation}
Clearly, the susceptibility function will depend upon the position of the
atom because of the Gaussian mode structure of the field, which will make a
difference with the plane-wave model of reference \cite{Hilico}.

Before considering this problem of position-dependence of the susceptibility
function, we come to characterize the second fluctuating term $d\left(
\delta {\cal E}_\alpha ^{{\rm se}}\right) $ appearing in equation (\ref
{deltaE}), which represents the spontaneous emission noise. Assuming a
dilute atomic medium such that the fluctuations of the different dipoles are
independent of each other, we obtain field fluctuations by summing up the
contributions of all dipoles:

\begin{equation}
\left\langle d\left( \delta {\cal E}_\alpha ^{{\rm se}}\left( {\bf r}%
,\left[ \omega \right] \right) \right) d\left( \delta {\cal E}^{\beta \ 
{\rm se}}\left( {\bf r}^{\prime },\left[ \omega ^{\prime }\right]
\right) \right) \right\rangle =2\pi \delta \left( \omega +\omega ^{\prime
}\right) \delta \left( {\bf r}-{\bf r}^{\prime }\right) \frac{3\lambda
^2}{4\pi }\rho \left( {\bf r}\right) dx\ \sigma _\alpha ^\beta \left( 
{\bf r},\left[ \omega \right] \right)
\end{equation}
The correlation functions $\sigma _\alpha ^\beta $ describe the fluctuations
of the dipole components evaluated for one atom at point ${\bf r}$:

\begin{equation}
\sigma _\alpha ^\beta \left( {\bf r},\tau \right) =\left\langle \delta 
{\cal S}_\alpha \left( \tau \right) \delta {\cal S}^\beta \left(
0\right) \right\rangle _{{\bf r}}  \label{sigmat}
\end{equation}
The analytical expressions of the Fourier transforms $\sigma _\alpha ^\beta
\left( {\bf r},\left[ \omega \right] \right) $, as well as those of $\chi
_\alpha ^\beta \left( {\bf r},\left[ \omega \right] \right) $, can be
obtained in terms of a single atomic function $G_\alpha ^\beta $ which will
be given in the next section.

As already carried out for the mean values, we now solve the problem of the
transverse dependence of correlation functions by supposing that the
fluctuations radiated by the atoms into the higher-order modes have a
negligible effect on intracavity field fluctuations, and furthermore are not
matched to the detection setup. We will therefore obtain the modification $%
d\left( \delta {\cal A}_\alpha \right) $ of the amplitude fluctuations
due to the atomic medium by projecting the field fluctuations $d\left(
\delta {\cal E}_\alpha \right) $ onto the Gaussian mode functions:

\[
d\left( \delta {\cal A}_\alpha \left( t\right) \right) =\int d\left(
\delta {\cal E}_\alpha \left( {\bf r},t\right) \right) {\cal U}%
_\alpha ^{*}\left( {\bf r}\right) dy\ dz 
\]
Considering first the linear response term, we project the effect of a thin
atomic layer of length $dx$ (cf. equation (\ref{deltaElr})) and sum up over
the whole atomic medium. We thus write $d\left( \delta {\cal A}_\alpha ^{%
{\rm lr}}\right) $ in the frequency domain in terms of a new susceptibility
function $\overline{\chi }_\alpha ^\beta \left[ \omega \right] $:

\begin{equation}
d\left( \delta {\cal A}_\alpha ^{{\rm lr}}\left[ \omega \right] \right)
=-\frac{3\lambda ^2}{4\pi S}n\displaystyle{\sum_{\beta}}\overline{\chi }%
_\alpha ^\beta \left[ \omega \right] \delta {\cal A}_\beta \left[ \omega
\right]
\end{equation}
which is defined as the local susceptibility function averaged over the beam
profile:

\begin{equation}
\overline{\chi }_\alpha ^\beta \left[ \omega \right] =\frac 1n\int \chi
_\alpha ^\beta \left( {\bf r},\left[ \omega \right] \right) \rho \left( 
{\bf r}\right) f\left( {\bf r}\right) e^{i\left( \varepsilon _\alpha
-\varepsilon _\beta \right) \varphi \left( {\bf r}\right) }d{\bf r}
\label{khib}
\end{equation}

For the spontaneous emission noise, we once more disregard the fluctuations
radiated by the atoms into the higher-order modes, then project the field
fluctuations $d\delta {\cal E}_\alpha ^{{\rm se}}$ onto the Gaussian
mode, and finally sum up the independent contributions of all atoms. We thus
obtain the contribution of spontaneous emission noise in terms of a
correlation function $\overline{\sigma }_\alpha ^\beta \left[ \omega \right] 
$ averaged over the transverse profile of the beam:

\begin{equation}
\left\langle d\left( \delta {\cal A}_\alpha ^{{\rm se}}\left[ \omega
\right] \right) d\left( \delta {\cal A}^{\beta \ {\rm se}}\left[ \omega
^{\prime }\right] \right) \right\rangle =2\pi \delta \left( \omega +\omega
^{\prime }\right) \frac{3\lambda ^2}{4\pi S}n\overline{\sigma }_\alpha
^\beta \left[ \omega \right]
\end{equation}
\begin{equation}
\overline{\sigma }_\alpha ^\beta \left[ \omega \right] =\frac 1n\int \sigma
_\alpha ^\beta \left( {\bf r},\left[ \omega \right] \right) \rho \left( 
{\bf r}\right) f\left( {\bf r}\right) e^{i\left( \varepsilon _\alpha
-\varepsilon _\beta \right) \varphi \left( {\bf r}\right) }d{\bf r}
\label{sigmab}
\end{equation}

Now, the computation of quantum fluctuations in the output field proceeds
along the same lines as in the plane-wave model \cite{Hilico}, provided that
the following modifications are accounted for. First, field fluctuations are
understood as refering to the Gaussian mode amplitudes, for the intracavity
field $\delta {\cal A}_\alpha $ as well as for input and output fields $%
\delta {\cal A}_\alpha ^{{\rm in}}$ and $\delta {\cal A}_\alpha ^{%
{\rm out}}$. Then, the plane-wave susceptibility function $\chi _\alpha
^\beta $ and noise spectrum $\sigma _\alpha ^\beta $ are replaced by the
averaged expressions $\overline{\chi }_\alpha ^\beta $ and $\overline{\sigma 
}_\alpha ^\beta $ given respectively by equations (\ref{khib}) and (\ref
{sigmab}). This leads to a simple and natural generalisation of the results
of the plane-wave model. When transverse beam structure is accounted for,
the position-dependent expressions for atomic susceptibility function and
atomic noise spectrum have to be averaged over the beam intensity profile.
The same property was in fact used in published calculations accounting for
Gaussian profile of the laser beam \cite{Xiao87a,Hope90}: in these papers
where quantum stochastic methods were used, the squeezing spectra were
obtained by averaging the drift and diffusion matrices which describe in
these methods the effect of atoms. In the present paper, the fact that the
contributions of elementary volumes have to be added independently has not
been postulated. We have shown that this averaging property follows from a
simple treatment of Gaussian profile where fields radiated by the atoms are
projected onto the fundamental Gaussian mode.

\section{Evaluation of the averaged atomic spectra}

Before coming to the estimation of the squeezing spectra, we now give the
expressions of the atomic spectra and deduce the expressions averaged over
the intensity profile. We then compare the atomic spectra thus obtained with
those derived in the plane-wave model.

The local susceptibility function and noise spectrum $\chi _\alpha ^\beta
\left( {\bf r},\left[ \omega \right] \right) $ and $\sigma _\alpha ^\beta
\left( {\bf r},\left[ \omega \right] \right) $ are obtained by solving
optical Bloch equations for an atom located at point ${\bf r}$. They have
the same form as in the plane-wave model \cite{Reid88,Hilico}, with a
position-dependent field parameter $\beta \left( {\bf r}\right) $
however, and may be written in terms of the phase $\varphi \left( {\bf r}%
\right) $ of the Gaussian mode and of a function $G_\alpha ^\beta $ which
only depends on the local intensity parameter:

\begin{eqnarray}
\chi _\alpha ^\beta \left( {\bf r},\left[ \omega \right] \right) &=&\frac
i{2\gamma }\left( G_\alpha ^\beta \left( {\bf r},\left[ \omega \right]
\right) -G_{3-\alpha }^{3-\beta }\left( {\bf r},\left[ -\omega \right]
\right) ^{*}\right) e^{-i\left( \varepsilon _\alpha -\varepsilon _\beta
\right) \varphi \left( {\bf r}\right) }  \label{khiPW} \\
\sigma _\alpha ^\beta \left( {\bf r},\left[ \omega \right] \right)
&=&\frac 1{2\gamma }\left( G_\alpha ^\beta \left( {\bf r},\left[ \omega
\right] \right) +G_\beta ^\alpha \left( {\bf r},\left[ \omega \right]
\right) ^{*}\right) e^{-i\left( \varepsilon _\alpha -\varepsilon _\beta
\right) \varphi \left( {\bf r}\right) }  \label{sigmaPW}
\end{eqnarray}
\begin{equation}
G_\alpha ^\beta \left( {\bf r},\left[ \omega \right] \right) =\frac{%
p_\alpha ^\beta \left( \left| \beta \left( {\bf r}\right) \right|
^2,\left[ \omega \right] \right) }{q\left( \left| \beta \left( {\bf r}%
\right) \right| ^2,\left[ \omega \right] \right) }  \label{Gomega}
\end{equation}
where:

\begin{eqnarray}
p_1^1\left( \left| \beta \right| ^2,\left[ \omega \right] \right) &=&\left(
1+\delta ^2+2\left| \beta \right| ^2\right) \left( 1+\delta ^2\right) \left(
1-i\delta -i\overline{\omega }\right) \left( 2-i\overline{\omega }\right) 
\nonumber \\
&&+2\left| \beta \right| ^2\left( 1+\delta ^2+2\left| \beta \right|
^2\right) i\overline{\omega }\left( 1+i\delta \right)  \nonumber \\
&&+2\left| \beta \right| ^4\left( 2-i\overline{\omega }\right) ^2+4\left|
\beta \right| ^6 \\
p_2^1\left( \left| \beta \right| ^2,\left[ \omega \right] \right) &=&2\left|
\beta \right| ^2\left( 1+\delta ^2+2\left| \beta \right| ^2\right) \left( 2-i%
\overline{\omega }\right) \left( 1+i\delta \right)  \nonumber \\
&&+\left| \beta \right| ^2\left( 1+i\delta \right) ^2\left( 1+i\delta -i%
\overline{\omega }\right) \left( 2-i\overline{\omega }\right)  \nonumber \\
&&+2\left| \beta \right| ^4i\overline{\omega }\left( 1+i\delta \right)
+4\left| \beta \right| ^6 \\
p_1^2\left( \left| \beta \right| ^2,\left[ \omega \right] \right) &=&\left|
\beta \right| ^2\left( 1-i\delta \right) ^2\left( 1-i\delta -i\overline{%
\omega }\right) \left( 2-i\overline{\omega }\right)  \nonumber \\
&&+2\left| \beta \right| ^4i\overline{\omega }\left( 1-i\delta \right)
+4\left| \beta \right| ^6 \\
p_2^2\left( \left| \beta \right| ^2,\left[ \omega \right] \right) &=&2\left|
\beta \right| ^4\left( 2-i\overline{\omega }\right) ^2+4\left| \beta \right|
^6 \\
q\left( \left| \beta \right| ^2,\left[ \omega \right] \right) &=&\left(
1+\delta ^2+2\left| \beta \right| ^2\right) ^2\left( 2-i\overline{\omega }%
\right) \left( \left( 1-i\overline{\omega }\right) ^2+\delta ^2\right) 
\nonumber \\
&&+\left( 1+\delta ^2+2\left| \beta \right| ^2\right) ^24\left| \beta
\right| ^2\left( 1-i\overline{\omega }\right)
\end{eqnarray}
$\delta $ and $\beta $ are the dimensionless parameters already defined for
measuring laser-atom detuning and Rabi frequency respectively; $\overline{%
\omega }$ is the noise frequency also normalized to the decay rate of the
dipole:

\begin{equation}
\overline{\omega }=\frac \omega \gamma
\end{equation}

In order to perform the average over the beam profile, we first note that
the phases do not enter the averaged expressions (compare equations (\ref
{khib},\ref{sigmab}) with equations (\ref{khiPW},\ref{sigmaPW})). We then
proceed by analogy with the computation of the factors $n^{(1+k)}$ which
appeared in the expression (\ref{powerexpansionN}) of the effective number
of atoms, by developing the functions $G_\alpha ^\beta $ in terms of the
intensity parameter:

\begin{equation}
G_\alpha ^\beta \left( {\bf r},\left[ \omega \right] \right) =
\displaystyle{\sum_{k=0}^{\infty}} G_\alpha ^{\beta \ \left( k\right) }\left[
\omega \right] \left| \beta \left( {\bf r}\right) \right| ^{2k}
\end{equation}
Writing the position-dependent local intensity as in the previous discussion
of the bistability curve:

\begin{equation}
\left| \beta \left( {\bf r}\right) \right| ^2=\left| \beta _{{\rm m}%
}\right| ^2f\left( {\bf r}\right) =2Xf\left( {\bf r}\right)
\end{equation}
we obtain:

\begin{eqnarray}
\overline{\chi }_\alpha ^\beta \left[ \omega \right] &=&
\displaystyle{\sum_{k=0}^{\infty}}\frac{i\xi _kX^k}{2\gamma }\left( G_\alpha ^{\beta
\ \left( k\right) }\left[ \omega \right] -G_{3-\alpha }^{3-\beta \ \left(
k\right) }\left[ -\omega \right] ^{*}\right)  \label{khiGW} \\
\overline{\sigma }_\alpha ^\beta \left[ \omega \right] &=&
\displaystyle{\sum_{k=0}^{\infty}}\frac{\xi _kX^k}{2\gamma }\left( G_\alpha ^\beta
\left[ \omega \right] +G_\beta ^{\alpha \ \left( k\right) }\left[ \omega
\right] ^{*}\right)  \label{sigmaGW}
\end{eqnarray}
where:

\begin{equation}
\xi _k=\frac{2^k}{1+k}
\end{equation}

The plane-wave expressions $\chi _\alpha ^\beta \left[ \omega \right] $ and $%
\sigma _\alpha ^\beta \left[ \omega \right] $ written in equations (\ref
{khiPW},\ref{sigmaPW}) may be identified with equations (\ref{khiPW},\ref
{sigmaPW}) with $\xi _k$ replaced by unity (global phase factors are
unessential and may be disregarded). The definitions of $X$, standing for
the position-independent intensity parameter $\left| \beta \right| ^2$ in
the plane-wave model and for its spatial average over the beam profile in
the Gaussian-wave theory, have been precisely chosen so that the
coefficients $\xi _0$ and $\xi _1$ are equal to unity. This implies that the
differences between the plane-wave and Gaussian-beam predictions for
squeezing only come from the fact that the higher-order coefficients $\xi _k$
(with $k=2,3\ldots $) differ from unity. The discussion of these differences
may therefore be restricted to the analysis of the contributions to atomic
spectra proportional to $X^2$, $X^3\ldots $: the plane-wave predictions
cannot be considered as reliable as soon as these contributions become
appreciable.

It could be expected that the contributions proportional to $X^2$, $%
X^3\ldots $ to the atomic spectra become appreciable, as for the mean
values, only when the excited state becomes significantly populated, that is
when the intensity parameter $X$ approaches $\delta ^2$. The plane-wave
predictions would thus be reliable in the whole range $\delta <X<\delta ^2$
where a large squeezing is expected (see the discussion at the end of
section 3). We now show that this is not the case.

We first consider the limiting case, that we will call the Kerr domain in
the following, where only the terms proportional to $X^0$ and $X^1$ are
appreciable. In this case, the coincidence between the atomic spectra
obtained in the two models is guaranteed. The linear terms (proportional to $%
X^0$) are given by:

\begin{eqnarray}
G_1^{1\ \left( 0\right) } &=&\frac 1{1+i\delta -i\overline{\omega }} \\
G_2^{1\ \left( 0\right) } &=&G_1^{2\ \left( 0\right) }=G_2^{2\ \left(
0\right) }=0
\end{eqnarray}
and the Kerr terms (proportional to $X^1$) by: 
\begin{eqnarray}
G_1^{1\ \left( 1\right) } &=&\frac{-2\left( 2+2i\delta -i\overline{\omega }%
\right) }{\left( 1+\delta ^2\right) \left( 1+i\delta -i\overline{\omega }%
\right) ^2} \\
G_2^{1\ \left( 1\right) } &=&\frac 2{\left( 1-i\delta \right) \left( \left(
1-i\overline{\omega }\right) ^2+\delta ^2\right) }+\frac 1{\left( 1-i\delta
\right) ^2\left( 1-i\delta -i\overline{\omega }\right) } \\
G_1^{2\ \left( 1\right) } &=&\frac 1{\left( 1+i\delta \right) ^2\left(
1+i\delta -i\overline{\omega }\right) } \\
G_2^{2\ \left( 1\right) } &=&0
\end{eqnarray}
These terms have a simple physical interpretation in terms of a diagrammatic
perturbative expansion of the field-matter interaction \cite
{Dalibard83,Heidmann87}. The linear terms are described by two-photon
elastic scattering amplitudes which are resonant at frequencies close to the
atomic eigenfrequency ($\overline{\omega }\simeq \delta $) while the Kerr
terms are described by four-photon inelastic scattering amplitudes which are
resonant at frequencies either close to the atomic eigenfrequency ($%
\overline{\omega }\simeq \delta $) or symmetric to the atomic eigenfrequency
with respect to the laser frequency ($\overline{\omega }\simeq -\delta $).
The interpretation of squeezing in terms of these photon scattering
amplitudes was already performed for the case of atoms coupled to field in
free space \cite{Heidmann85,Heidmann87}. The previous equations give a more
complete description of the effect of these amplitudes which may be used
also for interpreting squeezing by atoms coupled to field in a cavity. An
important consequence of this discussion is that the squeezing processing is
resonant on the sidebands of fluorescence triplet ($\overline{\omega }\simeq
\pm \delta $) and not on the central component ($\overline{\omega }\simeq 0$%
). In contrast, squeezing via bistability is usually studied for frequencies
close to the laser frequency. This is particularly clear in the good-cavity
limit where the entire cavity bandwidth is contained in the central
component of the emission triplet. In this case, the previous equations have
to be evaluated at zero frequency ($\overline{\omega }=0$) where they
correspond to non-resonant processes.

We come now to the analysis of the higher-order contributions to atomic
spectra proportional to $X^2$, $X^3\ldots $, and focus the discussion on the
same regime where squeezing is studied for frequencies close to the laser
frequency. In this regime, it is easily checked on the expressions of the
atomic susceptibility functions $\chi _\alpha ^\beta $ that the higher-order
contributions remain negligible as long as $X\ll \delta ^2$. This is related
to the general property of linear response theory which states that static
susceptibilities evaluated at $\overline{\omega }=0$ are directly related to
differentiated forms of the relations between mean fields, and to the
already known fact that the plane-wave model provides reliable expressions
for mean fields when $X\ll \delta ^2$. However, the higher-order
contributions to the noise spectra $\sigma _\alpha ^\beta $ may become
appreciable even when $X\ll \delta ^2$. This can be shown by considering the
atomic spectra at zero frequency (obtained by setting $\overline{\omega }=0$
in the unapproximated forms (\ref{Gomega})):

\begin{eqnarray}
G_1^1\left[ 0\right] &=&\frac{\left( 1+\delta ^2\right) \left( 1-i\delta
\right) }{\left( 1+\delta ^2+2\left| \beta \right| ^2\right) ^2}+\frac{%
4\left| \beta \right| ^4+2\left| \beta \right| ^6}{\left( 1+\delta
^2+2\left| \beta \right| ^2\right) ^3} \\
G_2^1\left[ 0\right] &=&\frac{2\left| \beta \right| ^2\left( 1+i\delta
\right) }{\left( 1+\delta ^2+2\left| \beta \right| ^2\right) ^2}+\frac{%
\left| \beta \right| ^2\left( 1+i\delta \right) ^3+2\left| \beta \right| ^6}{%
\left( 1+\delta ^2+2\left| \beta \right| ^2\right) ^3} \\
G_1^2\left[ 0\right] &=&\frac{\left| \beta \right| ^2\left( 1-i\delta
\right) ^3+2\left| \beta \right| ^6}{\left( 1+\delta ^2+2\left| \beta
\right| ^2\right) ^3} \\
G_2^2\left[ 0\right] &=&\frac{4\left| \beta \right| ^4+2\left| \beta \right|
^6}{\left( 1+\delta ^2+2\left| \beta \right| ^2\right) ^3}
\end{eqnarray}
The evaluation of the susceptibility spectra relies only upon the first
terms appearing in $G_1^1\left[ 0\right] $ and $G_2^1\left[ 0\right] $, so
that the higher-order contributions to these terms are negligible when $X\ll
\delta ^2$. In contrast, the other terms, which enter in the evaluation of
the noise spectra, may be appreciable even for $X\ll \delta ^2$. Restricting
our interest to this low-saturation limit, we may identify the main
contribution to the squeezing process as the parametric terms proportional
to $X/\delta ^3$ which appear in $G_2^1\left[ 0\right] $. Among the dominant
noise contributions, there are the terms proportional to $1/\delta ^2$ in $%
G_1^1\left[ 0\right] $ and $X/\delta ^4$ in $G_2^1\left[ 0\right] $ or $%
G_1^2\left[ 0\right] $, but also the higher-order terms proportional to $%
X^2/\delta ^6$ or $X^3/\delta ^6$ which appear in the four functions $%
G_\beta ^\alpha \left[ 0\right] $. It is thus clear that these terms may not
be disregarded for $X\ll \delta ^2$. In particular, the noise term
proportional to $X^3/\delta ^6$ reaches the magnitude of the parametric Kerr
coefficient $X/\delta ^3$ for $X\approx \delta ^{3/2}$.

This noise term is not accounted for in a Kerr model and is found to be
responsible for excess noise and degradation of the expected squeezing \cite
{Reid85}. From the qualitative evaluations of previous paragraph, one
deduces that the noise correlation functions have to be suspected in the
domain $\delta <X<\delta ^2$, where a large squeezing is expected from
plane-wave computations (see conditions (\ref{conditionC},\ref{conditionX}%
)). This does not imply that plane-wave predictions of squeezing are wrong
in the whole domain, but this forces us to perform the Gaussian-wave
computations in order to get reliable expectations. In the next section, we
give a quantitative estimate of squeezing degradation due to this effect and
delineate the domain of validity of plane-wave computations.

The conclusion of this discussion is that, although the plane-wave model
provides reliable estimations for the bistability curve and for the
susceptibility functions in the whole low-saturation domain $X\ll \delta ^2$%
, this is not the case for the noise correlation functions. A qualitative
interpretation of this result may be found by coming back to photon
scattering amplitudes. As already discussed, the two-photon and four-photon
amplitudes are resonant on the sidebands of the fluorescence triplet. In
contrast, there exists higher-order scattering amplitudes which give rise to
inelastic fluorescence on the central component of the triplet \cite
{Dalibard83}. Although these amplitudes are proportional to higher-order
powers of $X$, they are favored with respect to the lowest-order amplitudes,
due to their resonant enhancement around the central frequency. This
explains why they can have an influence in the low-saturation domain.

\section{Squeezing spectra}

We come now to the computation of quantum fluctuations in the output field.
As already discussed, this computation proceeds along the same lines in the
Gaussian-wave computations as in the plane-wave model, provided that the
matrices $\chi _\alpha ^\beta $ and $\sigma _\alpha ^\beta $ are replaced by
the averaged expressions $\overline{\chi }_\alpha ^\beta $ and $\overline{%
\sigma }_\alpha ^\beta $.

The covariance functions for the input or output fields are defined in terms
of noise matrices:

\begin{eqnarray}
\left\langle \delta {\cal A}_\alpha ^{{\rm in}}\left[ \omega \right]
\delta {\cal A}^{\beta \ {\rm in}}\left[ \omega ^{\prime }\right]
\right\rangle &=&2\pi \delta \left( \omega +\omega ^{\prime }\right) 
{\cal V}_\alpha ^{\beta \ {\rm in}}\left[ \omega \right] \\
\left\langle \delta {\cal A}_\alpha ^{{\rm out}}\left[ \omega \right]
\delta {\cal A}^{\beta \ {\rm out}}\left[ \omega ^{\prime }\right]
\right\rangle &=&2\pi \delta \left( \omega +\omega ^{\prime }\right) 
{\cal V}_\alpha ^{\beta \ {\rm out}}\left[ \omega \right]
\end{eqnarray}
${\cal V}_\alpha ^{\beta \ {\rm in}}$ and ${\cal V}_\alpha ^{\beta \ 
{\rm out}}$ are the elements of square matrices ${\cal V}^{{\rm in}}$
and ${\cal V}^{{\rm out}}$ which are connected through noise processing
relations (equations (47-51) of reference \cite{Hilico}):

\begin{equation}
{\cal V}^{{\rm out}}\left[ \omega \right] =\left( \mu \left[ \omega
\right] -1\right) .{\cal V}^{{\rm in}}\left[ \omega \right] .\left( \mu
\left[ \omega \right] -1\right) ^{\dagger }+\mu \left[ \omega \right] .%
{\cal V}^{{\rm at}}\left[ \omega \right] .\mu \left[ \omega \right]
^{\dagger }  \label{Vout}
\end{equation}
$\left( \mu \left[ \omega \right] -1\right) $ and $\mu \left[ \omega \right] 
$ are transfer matrices which characterize noise processing by the cavity
containing atoms and which depends on the susceptibility matrix $\overline{%
\chi }$:

\begin{equation}
\mu \left[ \omega \right] =\frac 2{1+i\phi -i\frac \omega \kappa -2\gamma
C\varepsilon .\overline{\chi }\left[ \omega \right] }
\end{equation}
${\cal V}^{{\rm at}}$ describes the noise added by the atomic medium,
and which depends on the atomic noise matrix $\overline{\sigma }$:

\begin{equation}
{\cal V}^{{\rm at}}\left[ \omega \right] =\gamma C\varepsilon .\overline{%
\sigma }\left[ \omega \right] .\varepsilon
\end{equation}
In these expressions, $\varepsilon $ is a diagonal matrix containing the
coefficients $\varepsilon _\alpha $:

\begin{equation}
\varepsilon =\left[ 
\begin{tabular}{ll}
1 & 0 \\ 
0 & -1
\end{tabular}
\right]
\end{equation}
It is worth stressing at this stage that, while the atomic spectra $%
\overline{\chi }$ and $\overline{\sigma }$ are obtained by averaging the
plane-wave spectra over the intensity profile, this is not the case for the
noise spectra of the output fields.

We may now give a quantitative estimation of the influence of the Gaussian
profile upon the noise spectra for the output fields. The transfer matrices
are not modified in the low-saturation limit, since they depend only upon
the susceptibility matrix $\overline{\chi }$. It follows that the first term
in the right-hand side of equation (\ref{Vout}) is not influenced by the
Gaussian profile. In contrast, the second term, which is proportional to the
atomic noise matrix $\overline{\sigma }$, has a larger degrading effect on
squeezing in the Gaussian-wave theory than in the plane-wave model.
Precisely, the squeezing degradation due to higher-order noise terms is
greater in Gaussian-wave theory than in the plane-wave model; for terms
proportional to $X^2$ (respectively $X^3$), this degradation is multiplied
by $\xi _2=4/3$ (respectively by $\xi _3=2$). In the Kerr domain, those
higher-order noise terms have a negligible influence on squeezing, so that
the predictions of plane-wave model coincide with those of Gaussian-wave
theory.

A more detailed discussion may be given by studying the example of the
optimum squeezing expected at zero frequency ($\omega =0$) on a bistability
turning point (defined by equations (\ref{condzeta},\ref{condphi})). The
optimum squeezing (variance of the quadrature component with minimal
fluctuations, evaluated at zero frequency on a bistability turning point)
reaches in the Kerr domain the value:

\begin{equation}
S_{{\rm opt}}^{{\rm Kerr}}=\frac a{1+a}
\end{equation}
This value is small, of the order of $2C/\delta ^2$, when condition (\ref
{conditionC}) is obeyed, that is when the effect of atomic absorption on the
field remains small. Note that a small value of $S_{{\rm opt}}$ means an
efficient squeezing of input field fluctuations by the bistable cavity.
Assuming that the intensity parameter $X$ lies in the domain $\delta \ll
X\ll \delta ^2$, a close inspection of the equations describing noise
processing by the bistable cavity shows that the optimum squeezing $S_{{\rm %
opt}}$ accounting for higher-order noise terms in the plane-wave model may
be written:

\begin{equation}
S_{{\rm opt}}\simeq \frac{S_{{\rm opt}}^{{\rm Kerr}}+2f\frac{X^2}{\delta
^3}}{1+2f\frac{X^2}{\delta ^3}}  \label{sopt}
\end{equation}
The numerical factor $f$ has the following value, of the order of unity ($%
\phi $ is defined by (\ref{condphi})):

\begin{equation}
f=\frac{\zeta -\left( \phi -a\delta \right) }{1+a}
\end{equation}
The degradation due to higher-order noise terms is thus found to become
appreciable when $2X^2/\delta ^3\simeq 2C/\delta ^2$, that is when:

\begin{equation}
X\simeq \sqrt{C\delta }
\end{equation}
The degradation of squeezing appearing in equation (\ref{sopt}) is enlarged
when passing from the plane-wave to the Gaussian-wave theory. It turns out
that, in the limiting case where degradation is small, it may be attributed
mainly to higher-order atomic noise terms proportional to $X^3$, so that
squeezing degradation is roughly multiplied by $\xi _3=2$ due to Gaussian
transverse profile. In other words, the plane-wave model systematically
underestimates the degrading effect of higher-order contributions to atomic
noise and does not provide reliable estimations for the squeezing when such
contributions cannot be neglected.

In contrast, we may consider now the condition which delineates the Kerr
region:

\begin{equation}
X\ll \sqrt{C\delta }  \label{condKerr}
\end{equation}
When this condition is satisfied, the optimum squeezing is not degraded by
higher-order noise terms, so that the plane-wave computation provides a
reliable estimation for squeezing. It is worth recalling here that this
condition has been derived in the domain $\delta \ll X\ll \delta ^2$ where a
large squeezing is expected. In the domain of very low intensity $X<\delta $%
, different conclusions would be drawn for reliability of the plane-wave
model, but a poor squeezing would be predicted.

The previous discussion was restricted to the particular case of zero
frequency where analytical expressions may be handled more easily. Figure 1
shows spectra for the optimum squeezing evaluated as a function of noise
frequency $\omega $. 
\begin{figure}
\centerline{\psfig{figure=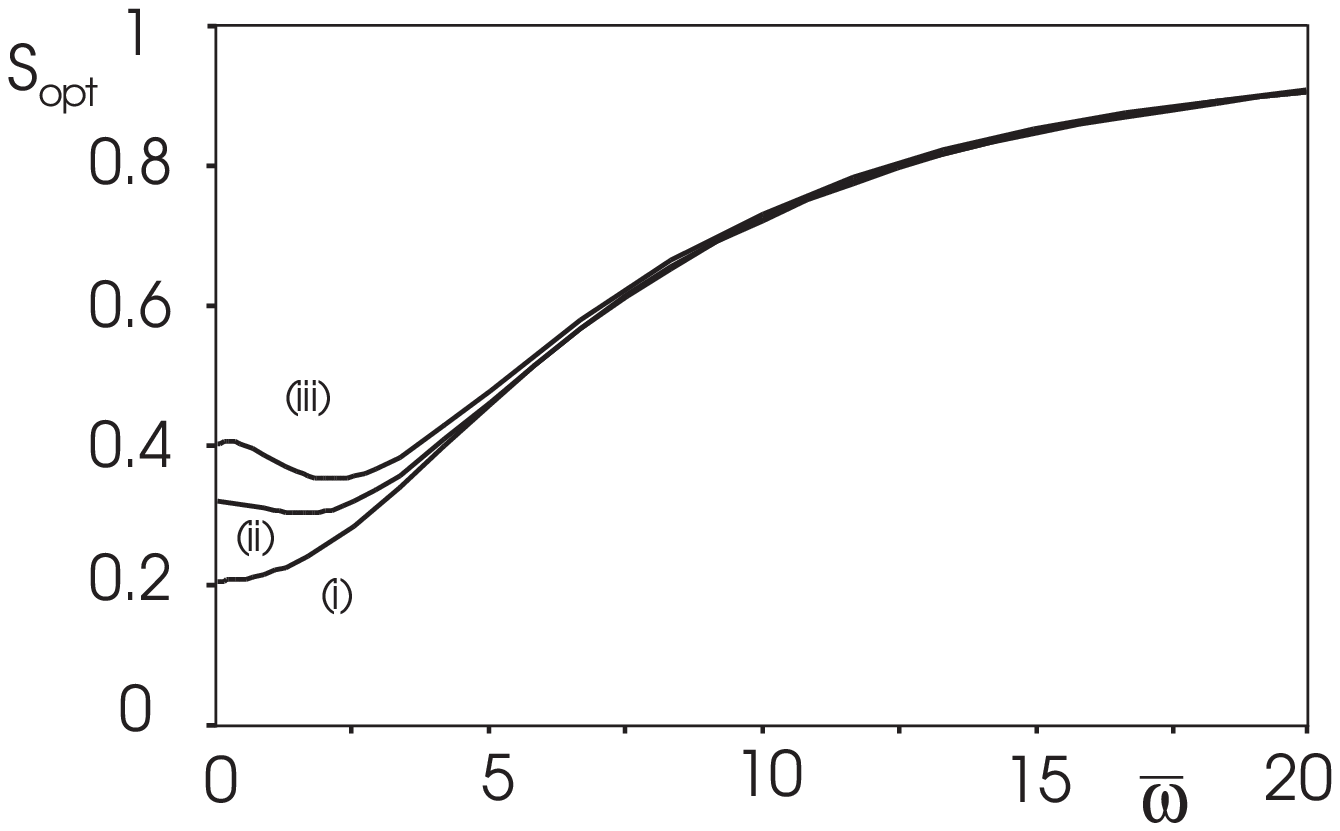,height=5cm}}
\caption{Spectra for the optimum squeezing as a function of
normalized noise frequency $\overline{\omega }=\omega /\gamma $, with a
decay constant for the intracavity field $\kappa =10\gamma $; the three
spectra correspond to the same atomic absorption coefficient ($a=0.25$) and
to the same turning point at bistability threshold ($\zeta =1+a$; $\phi
=a\delta $); the spectrum (i) is computed for a large detuning lying in the
Kerr domain ($\delta =1000$; the other parameters deduced from the values of
$\delta $, $a$ and $\zeta $ are $X=2500$, $C=126253$, $\phi =250$); the two
other spectra are computed for a lower detuning lying outside the Kerr
domain ($\delta =100$; $X=250$, $C=1378$, $\phi =25$) and result from
plane-wave computations (ii) and Gaussian-wave computations (iii).}
\end{figure}
\noindent The three spectra correspond to the same value for the
atomic absorption coefficient $a$ and to the same working point on the
bistability curve; we have chosen the bistability threshold such that there
is only one turning point, which corresponds to $\zeta =1+a$ and $\phi
=a\delta $ ($f=1$ in this particular case). The first spectrum (trace (i))
is computed for a detuning sufficiently large so that the Kerr limit is
reached. In this case, the spectra computed from plane-wave model and
Gaussian-wave theory cannot be distinguished. The two other spectra are
computed for a lower detuning, such that higher-order noise terms degrade
optimum squeezing. The results of plane-wave computations (trace (ii)) and
Gaussian-wave computations (trace (iii)) now differ, revealing in particular
a larger degradation when transverse structure is accounted for. In
addition, these spectra clearly show, in consistency with the discussion at
the end of section 5, that the excess noise responsible for squeezing
degradation is associated with the central peak of the atomic fluorescence
spectrum, whose width equals $2\gamma $, where $\gamma $ is the decay
constant of the atomic dipole, in the limiting case of low saturation \cite
{CCTSR78}. Note that the spectra of Figure 1 have been computed in the
bad-cavity limit where the decay constant $\kappa $ of the intracavity field
is larger than $\gamma $. In the opposite good-cavity limit, the resonant
behaviour of the higher-order excess noise would not be apparent on the
squeezing spectrum.

\section{Conclusion}

We have used the linear input-output formalism generalized to incorporate
the treatment of atomic fluctuations to discuss the effect of Gausian
transverse structure on squeezing with two-level atoms. We have shown that
this effect is described quite naturally as a spatial averaging of the
susceptibility and noise functions which appear in the linear input-output
formalism. This property has not been postulated a priori, but derived as a
consequence of the simple assumptions made in the treatment of transverse
profile: higher-order transverse modes have been supposed either far from
resonance, or of such a high order that they are not efficiently coupled to
the Gaussian mode.

The mean fields and the bistability curves deduced from plane-wave and
Gaussian-wave calculations coincide in the low-saturation limit where the
population of the atomic excited state remains small ($X\ll \delta ^2$). We
have shown that this is also the case for the atomic susceptibility
functions evaluated around zero frequency. This is related to the general
property of linear response theory that the static susceptibilities are
differentiated forms of the relations between mean fields. In contrast, the
limit of low saturation is not sufficient for ensuring the coincidence of
plane-wave and Gaussian-wave predictions for atomic noise functions. This is
due to the existence of higher-order noise terms, arising from the central
peak of the fluorescence spectrum. These terms are able to degrade the
optimum squeezing when $X$ reaches the value $\sqrt{C\delta }$. We have
shown that these terms have a larger degrading effect in the Gaussian-wave
calculations as in the plane-wave ones. This implies that plane-wave
predictions of a large squeezing are reliable only in the Kerr domain, that
is for $X\ll \sqrt{C\delta }$. This criterium for reliability of plane-wave
model appears quite different from those published in previous references 
\cite{Xiao87a,Xiao87b,Hope90}.

{\bf Acknowledgements}

Thanks are due to Claude Fabre and Elisabeth Giacobino for discussions.

\end{document}